\documentclass{article}

\usepackage[english]{babel}
\usepackage{listings}
\makeatletter
\usepackage{jlcode}
\lstdefinestyle{code}{
  language=Julia, 
  showstringspaces=false,
  keywordstyle=\color{blue},
  commentstyle=\color{gray},
  identifierstyle=\color[RGB]{0,0,0},
  columns=fullflexible,
  keepspaces=true
}
\lstnewenvironment{code}{\lstset{style=code}}{}

\usepackage[letterpaper,top=2cm,bottom=2cm,left=3cm,right=3cm,marginparwidth=1.75cm]{geometry}

\usepackage{amsmath}
\usepackage{graphicx}
\usepackage[colorlinks=true, allcolors=blue]{hyperref}

\title{Ai4EComponentLib.jl: A Component-base Model Library in Julia}
\author{Yuebao Yang$^{1}$, Jingyi Yang$^2$, Mingtao Li\footnote{Corresponding author}\\Xi'an Jiaotong University}

\begin{document}
\maketitle

\begin{abstract}
Ai4EComponentLib.jl\footnote{https://ai4energy.github.io/Ai4EComponentLib.jl/dev/}(Ai4EComponentLib) is a component-base model library based on Julia language, which relies on the differential equation solver DifferentialEquations.jl\cite{rackauckas2017differentialequations} and the symbolic modeling tool Modelingtoolkit.jl\cite{ma2021modelingtoolkit}. To handle problems in different physical domains, Ai4EComponentLib tries to build them with component-base model. Supported by a new generation of symbolic modeling tools, models built with Ai4EComponentLib are more flexible and scalable than models built with traditional tools like Modelica. This paper will introduce the instance and general modeling methods of Ai4EComponentLib model library.
\end{abstract}

\section{Motivation}

In the field of modeling and simulation, there are already many modeling tools, such as Modelica\cite{tiller2001introduction}, gProms\cite{asteasuain2001dynamic}, Simulink\cite{chaturvedi2017modeling} and so on. They play an important role in different fields, and these tools have become essential tools for scientific researchers. But that doesn't mean they are perfect. Taking the open source Modelica as an example, Modelica has its own compiler that can compile the modeling language into C language, which is a complete solution chain for simulation problems. Its disadvantage is that it is not easy to expand and it cannot generate solutions of optimization problems (parameter identification, optimal control problems), etc. The ModelingToolkit in the Julia ecosystem solves the problem of insufficient scalability. Based on the work of ModelingToolkit, Ai4EComponentLib has built some model libraries, hoping to take advantage of its high scalability in the future.

\section{Ai4EComponentLib System Design}

The method of Ai4EComponentLib design system is the same with the traditional component modeling idea. But due to the high scalability of ModelingToolkit, we will look at this problem with a higher level of abstraction. The core of component-base model is the process, and the changes in energy and matter in the process. Two key points when designing the system are: internal process and external connection. To design a system is to design their processes and connections.

\begin{figure}
\centering
\includegraphics[width=1.0\textwidth]{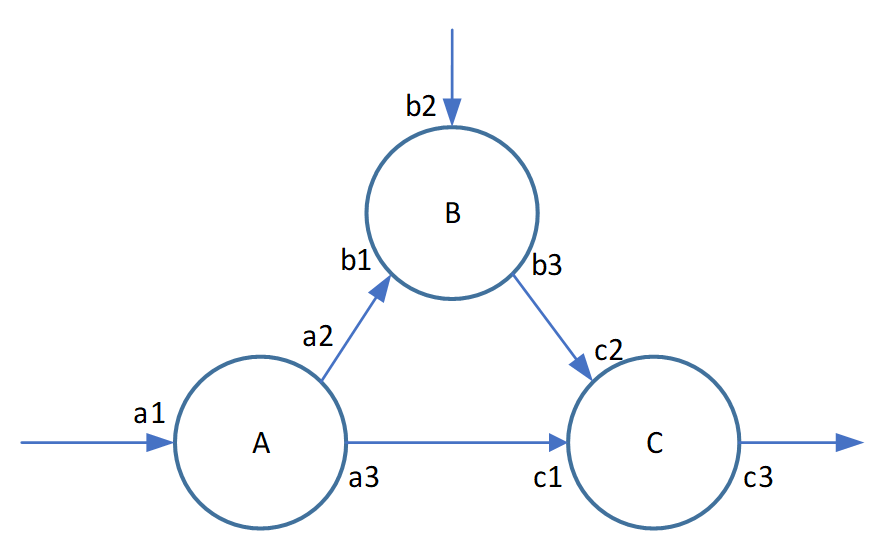}
\caption{\label{fig:f1}System with 3 components A, B and C.}
\end{figure}

In a system, there are 3 components A, B and C in the system. As shown in the Figure\ref{fig:f1}, each component has its own input and output nodes (such as a1, a2, etc.), and its inputs and outputs are shown by arrows.

\subsection{Internal Process}

Each component has its own characteristics, which is reflected in the process of the reaction between the input and output in the node or the exchange of material and energy. And it varies from component to component. These components are the mapping of real world to models and are abstractions of different physical phenomena.

For example, a1, a2, a3 of Component A in the above Figure\ref{fig:f1}. Inside the components, they have a specific mathematical relationship.

$$f(a1,a2,a3) = 0$$

Here, $f(a1,a2,a3)$ is a general functional form, which can be a differential equation or an algebraic equation. Take   a resistor component as an example, the same voltage difference is applied across a different resistor, and the current flowing through it is different. If the resistance changes with time, then the current of the system also changes with time. "Different resistance" and "time-varying resistance" are both characteristics of components, which are described by different equations $f$ when designing components. These equations determine what happens to "matter and energy" as it flows through the component. Once the components in the system are designed, the function of the whole system can be determined by these components.

\subsection{External connection}

When building system connections, we focus on the matter and energy of the connection points.

\begin{itemize}
\item If it is a circuit system. The a2 in Figure\ref{fig:f1} of the circuit system is used as the outlet of the component, and the current and voltage values are necessary attributes. And currents and voltage exist at the inlet and outlet of each component. So current and voltage are the connetor of the circuit system.
\item If it is a pipe system(without considering potential energy). The a2 in Figure\ref{fig:f1} of the pipe system is used as the outlet of the component, and the pressure and velocity of the water flow are necessary attributes. Water pressure and velocity properties exist at the inlet and outlet of each component. So pressure and velocity are the connetor of the pipe system.
\end{itemize}

The specific components needs to be build according to the physical model of the system. For example, Kirchhoff's laws in circuits and Bernoulli's equation for pipe system. The governing equations are the basic theory for designing the system.

In the external connection, the conservation law must generally be followed - the conservation of energy and the conservation of mass:

\[\left\{\begin{array}{l}
|a 2|=|b 1| \\ |b 3|=|c 2| \\|a 3|=|c 1|
\end{array}\right.\tag{1}
\]

For state variables, such as voltage, water pressure, gas pressure. Generally there are:

\[
\left\{\begin{array}{l}
a 2=b 1 \\ b 3=c 2 \\a 3=c 1
\end{array}\right.\tag{2}
\]

For process variables such as current, water flow, gas flow. Generally there are (specify that inflow is positive and outflow is negative):

\[
\left\{\begin{array}{l}
a 2+b 1=0 \\ b 3+c 2 =0\\a 3+c 1=0
\end{array}\right.\tag{3}
\]

\section{Case Study}

\subsection{Incompressible System}

When the liquid in the pipe is an incompressible fluid, the governing equation of the system is Bernoulli's equation\cite{munson2013fluid}.

\[\frac{p}{\rho g} +\frac{v^{2}}{2g}  + h=\text { constant } \tag{4}\]

Where, $p$ is pressure,$v$ is velocity, $h$ is height. Bernoulli's equation describes the law of conservation of energy for the flow of liquid inside a pipe. The internal process of the component is the increase or decrease of energy.

\subsubsection{Pipe Components}

For a straight pipeline, there are generally resistance losses along the route and local resistance losses inside the component. As the liquid flows from one port of the pipe to the other, the total energy of the liquid decreases. Along the way resistance loss and local resistance loss are internal processes of the pipeline assembly. Its equation is:

\[\frac{p_{in}}{\rho g} +\frac{8q_{in}^{2}}{2\pi^2D^4g} + z_{in}=
\frac{p_{out}}{\rho g} +\frac{8q_{out}^{2}}{2\pi^2D^4g} + z_{out}+h_f+h_m\tag{6}\]

Where the subscript $_{in}$ represents the pipeline inlet, $_{out}$ represents the pipeline outlet, ${h_f}$ represents the resistance loss along the way, and ${h_m}$ represents the local resistance loss.

\[h_f = f\frac{L}{D} \frac{8q^{2}}{\pi^2D^4g}\tag{7}\]

Where $f$ is the friction factor, $L$ is the pipe length, and $D$ is the pipe diameter.

\[h_m = K \frac{8q^{2}}{\pi^2D^4g}\tag{8}\]

Where $K$ is the local resistance coefficient.

The external connection of the pipeline is mainly pressure and flow (the height is not considered). Pressure is a state variable, and flow is a process variable. Between the two components a, b, there are:

\[p_a=p_b\\q_a+q_b=0\]

\subsubsection{Centrifugal Pump Components}

The internal process of a centrifugal pump is to add energy to the fluid.

\[\frac{p_{in}}{\rho g} +\frac{8q_{in}^{2}}{2\pi^2D^4g} + z_{in}=
\frac{p_{out}}{\rho g} +\frac{8q_{out}^{2}}{2\pi^2D^4g} + z_{out}+H_t\tag{9}\]

where $H_t$ represents the added energy when passing through the centrifugal pump.

\[H_t=\frac{(r\omega)^2}{g}-\frac{\omega \cot\beta }{2\pi bg}Q=c_0\omega^2-c_1\omega Q=a_0-a_1Q\tag{10}\]

Where $a_0, a_1$ represent the characteristic parameters of the pump, and Equation 10 is actually the theoretical head- flow curve of the pump.

\subsubsection{Components and Systems}

Before building the main components of the system such as pumps and pipes, we should build the Connector in ModelingToolkit (the variables in the Connector are pressure and flow in the pipeline system). Detailed code can be viewed in the code repository.

\begin{code}
# Component: SimplePipe(pipe with fixed friction factor `f`)
function SimplePipe(; name, L=10.0, D=25E-3, f=0.01, ρ=1.0E3, zin=0.0, zout=0.0, K_inside=0.0)
    @named in = PipeNode(z=zin)
    @named out = PipeNode(z=zout)
    ps = @parameters D = D L = L f = f K_inside = K_inside
    eqs = [
        _NodeEnergy(in, D, ρ) ~ _NodeEnergy(out, D, ρ) + _h_f(in, f, L, D) + _h_m(in, K_inside, D)
        0 ~ in.q + out.q
    ]
    compose(ODESystem(eqs, t, [], ps, name=name), in, out)
end

# Component: CentrifugalPump
function CentrifugalPump(; name, D=25E-3, ω=2500, c_0=4.4e-4, c_1=5.622, ρ=1.0E3)
    @named in = PipeNode()
    @named out = PipeNode()
    a_0 = c_0 * abs2(ω * 2π / 60)
    a_1 = c_1 * ω * 2π / 60
    ps = @parameters D = D
    eqs = [
        _NodeEnergy(in, D, ρ) + a_0 - a_1 * abs(in.q) ~ _NodeEnergy(out, D, ρ)
        0 ~ in.q + out.q
    ]
    compose(ODESystem(eqs, t, [], ps, name=name), in, out)
end

# Component: Sink_P
function Sink_P(; name, p=101325)
    @named port = PipeNode(z=0.0)
    eqs = [
        port.p ~ p
    ]
    compose(ODESystem(eqs, t, [], [], name=name), port)
end

\end{code}

\begin{figure}
\centering
\includegraphics[width=1.0\textwidth]{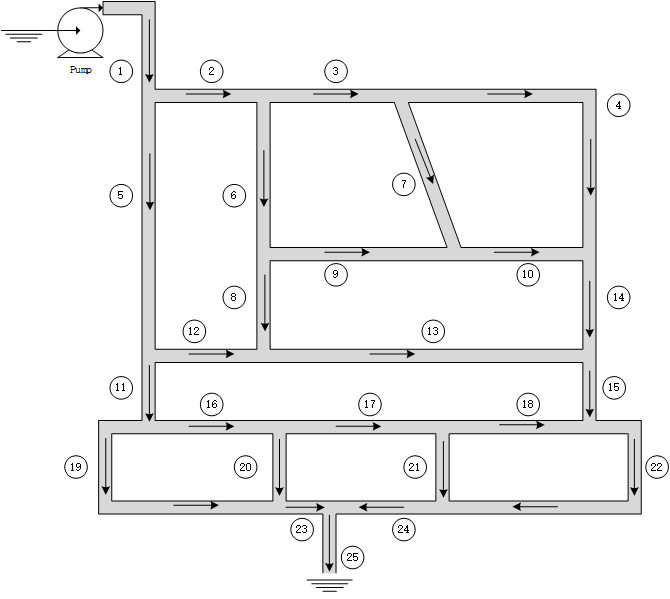}
\caption{\label{fig:f2}An Example of Pipe System}
\end{figure}

It becomes simple and fast to build the following system from the above components, as shown in Figure\ref{fig:f2}. 

\begin{code}

@named A = Sink_P()
@named B = Sink_P()

system = [A, B, Pump]

@named Pipe1 = SimplePipe(L=2.0);
push!(system, Pipe1);
@named Pipe2 = SimplePipe(L=3.0);
push!(system, Pipe2);
@named Pipe3 = SimplePipe(L=7.0);
push!(system, Pipe3);
@named Pipe4 = SimplePipe(L=9.0);
push!(system, Pipe4);
@named Pipe5 = SimplePipe(L=5.0);
push!(system, Pipe5);
@named Pipe6 = SimplePipe(L=4.0);
push!(system, Pipe6);
@named Pipe7 = SimplePipe(L=5.0);
push!(system, Pipe7);
@named Pipe8 = SimplePipe(L=1.0);
push!(system, Pipe8);
@named Pipe9 = SimplePipe(L=10.0);
push!(system, Pipe9);
@named Pipe10 = SimplePipe(L=2.0);
push!(system, Pipe10);
@named Pipe11 = SimplePipe(L=2.0);
push!(system, Pipe11);
@named Pipe12 = SimplePipe(L=3.0);
push!(system, Pipe12);
@named Pipe13 = SimplePipe(L=12.0);
push!(system, Pipe13);
@named Pipe14 = SimplePipe(L=1.0);
push!(system, Pipe14);
@named Pipe15 = SimplePipe(L=2.0);
push!(system, Pipe15);
@named Pipe16 = SimplePipe(L=3.0);
push!(system, Pipe16);
@named Pipe17 = SimplePipe(L=6.0);
push!(system, Pipe17);
@named Pipe18 = SimplePipe(L=6.0);
push!(system, Pipe18);
@named Pipe19 = SimplePipe(L=6.0);
push!(system, Pipe19);
@named Pipe20 = SimplePipe(L=1.0);
push!(system, Pipe20);
@named Pipe21 = SimplePipe(L=1.0);
push!(system, Pipe21);
@named Pipe22 = SimplePipe(L=7.0);
push!(system, Pipe22);
@named Pipe23 = SimplePipe(L=3.0);
push!(system, Pipe23);
@named Pipe24 = SimplePipe(L=3.0);
push!(system, Pipe24);
@named Pipe25 = SimplePipe(L=2.0);
push!(system, Pipe25);

eqs = [
  connect(A.port, Pump.in)
  connect(Pump.out, Pipe1.in)
  connect(Pipe1.out, Pipe2.in, Pipe5.in)
  connect(Pipe2.out, Pipe3.in, Pipe6.in)
  connect(Pipe3.out, Pipe4.in, Pipe7.in)
  connect(Pipe4.out, Pipe10.out, Pipe14.in)
  connect(Pipe5.out, Pipe11.in, Pipe12.in)
  connect(Pipe6.out, Pipe8.in, Pipe9.in)
  connect(Pipe7.out, Pipe9.out, Pipe10.in)
  connect(Pipe12.out, Pipe8.out, Pipe13.in)
  connect(Pipe13.out, Pipe14.out, Pipe15.in)
  connect(Pipe11.out, Pipe19.in, Pipe16.in)
  connect(Pipe16.out, Pipe17.in, Pipe20.in)
  connect(Pipe17.out, Pipe18.in, Pipe21.in)
  connect(Pipe18.out, Pipe15.out, Pipe22.in)
  connect(Pipe19.out, Pipe20.out, Pipe23.in)
  connect(Pipe21.out, Pipe22.out, Pipe24.in)
  connect(Pipe23.out, Pipe24.out, Pipe25.in)
  connect(B.port, Pipe25.out)
]
\end{code}

\subsection{Thermodynamic System}
\subsubsection{Math work}

In a thermodynamic cycle system, there are several typical internal processes\cite{moran2017engineering}.

\begin{itemize}
    \item Isothermal Process: $\frac{T}{P} = constant$
    \item Isobaric Process: $\frac{T}{v} = constant$
    \item Isentropic Process: $pv^k = constant$
    \item Isometric Process: $pv = constant$
\end{itemize}

The energy changes in these processes are ultimately reflected in the pressure, density, enthalpy, entropy, and temperature of the fluid. In the thermodynamic cycle, the external connection contains these 5 state variables.

When build a thermodynamic system, although governing equations for different internal processes can be used, there will always be some discrepancies between the theoretical equations and the real world. Therefore, when building a thermal system, we can take advantage of the scalability feature of ModelingToolkit—calling an external property library. Calling the property library can reduce the error of some theoretical calculations.

Taking an isothermal process as an example, the state of the starting point of the process is known. The governing equation of its internal process is:

\[T_1=T_2\tag{11}\]

Then determine another state variable at the end of the process, the other states at the end can be obtained from two known states. Then there are the following 4 combinations:

\[\begin{matrix}
 T_2,p_2 \Rightarrow  s_2,D_2,h_2\\
 T_2,s_2 \Rightarrow  p_2,D_2,h_2\\
 T_2,D_2 \Rightarrow  s_2,p_2,h_2\\
 T_2,h_2 \Rightarrow  s_2,p_2,D_2
\end{matrix}\tag{12}\]

Using the \textit{CoolProp} property library, enter the two parameters on the left side of the equation, and we can get any parameter on the right side of the equation. Therefore, the internal equations can be replaced by the external \textit{CoolProp} library.

\subsubsection{Components and Systems}

Taking the isenthalpic process as an example, the internal equation is:

$$h_1=h_2$$

If a parameter $p$ is given, the function \textit{choose\_equations} will automatically generate code that calls \textit{CoolProp} through $p,h$ to get the values of other parameters.

\begin{code}
function IsoenthalpyProcess(; name, inter_state="Q_0", fluid="Water")
    @assert inter_state != "H" "IsoenthalpyProcess can't accept H. Please chose another state."
    @named oneport = StreamPort()
    @unpack Δh, out = oneport
    eqs = [
        Δh ~ 0
    ]
    push!(eqs, chose_equations(out, inter_state, "H", fluid)...)
    return extend(ODESystem(eqs, t, [], []; name=name), oneport)
end
\end{code}

Consider the reheat cycle depicted in Figure\ref{fig:f3}:

\begin{figure}
\centering
\includegraphics[width=1.0\textwidth]{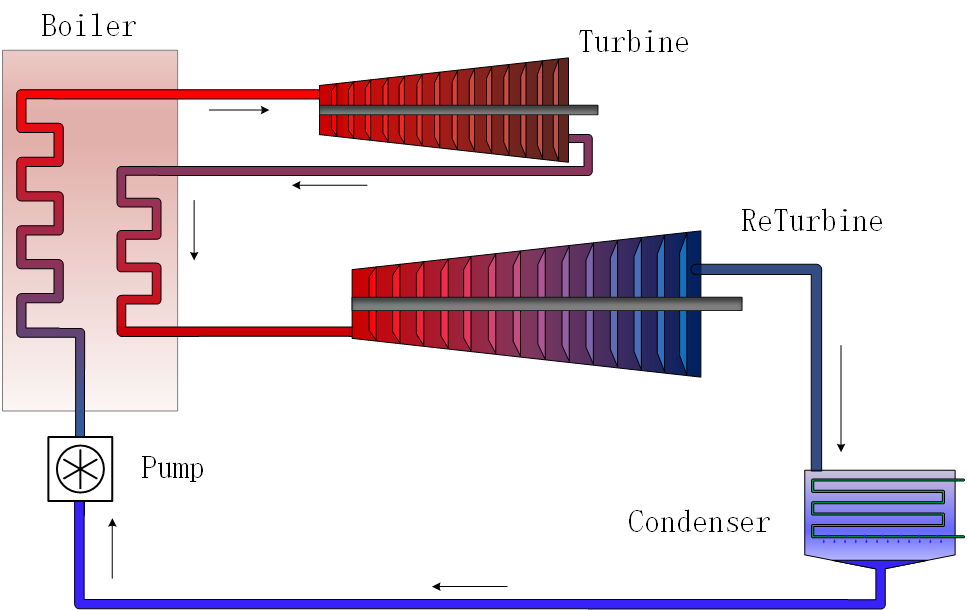}
\caption{\label{fig:f3}Reheat Rankine Cycle}
\end{figure}

The code is as follows. For more detailed, please check the code repository and documentation.

\begin{code}
@named pump = IsentropicProcess(inter_state="P")
@named pump_P = DThermalStates(state="P", value=-1.0E5, u0=18.0E6)

@named boiler = IsobaricProcess(inter_state="T")
@named boiler_T = ThermalStates(state="T", value=550+ 273.15)

@named turbine = IsentropicProcess(inter_state="P")
@named turbine_P = ThermalStates(state="P", value=3.0e6)

@named reboiler = IsobaricProcess(inter_state="T")
@named reboiler_T = ThermalStates(state="T", value=450 + 273.15)

@named returbine = IsentropicProcess(inter_state="P")
@named returbine_P = ThermalStates(state="P", value=4.0e3)

@named condenser = IsothermalProcess(inter_state="Q_0")

eqs = [
  connect(pump.out, boiler.in, pump_P.node)
  connect(boiler.out, turbine.in, boiler_T.node)
  connect(turbine.out, reboiler.in, turbine_P.node)
  connect(reboiler.out, returbine.in, reboiler_T.node)
  connect(returbine.out, condenser.in, returbine_P.node)
  connect(condenser.out, pump.in)
]
\end{code}

\section{Advantages of Component-base Models}

To summarize, components are essentially equations that describe the real world. The governing equations inside the component are the most important part, which lead the change of the system. External connections describe the relationship between components by connection equations. The connection equation does not affect the real system evolution, but only assists building the system.

Components-base model are designed to disassemble large systems into small components. The advantage is that it is convenient for users to build the system, and the disadvantage is that these connection variables will consume computing resources to be simplified after the system is built. In fact, the benefits of a componentized model far outweigh the cost of consuming a little computing resources.

\begin{figure}
\centering
\includegraphics[width=0.6\textwidth]{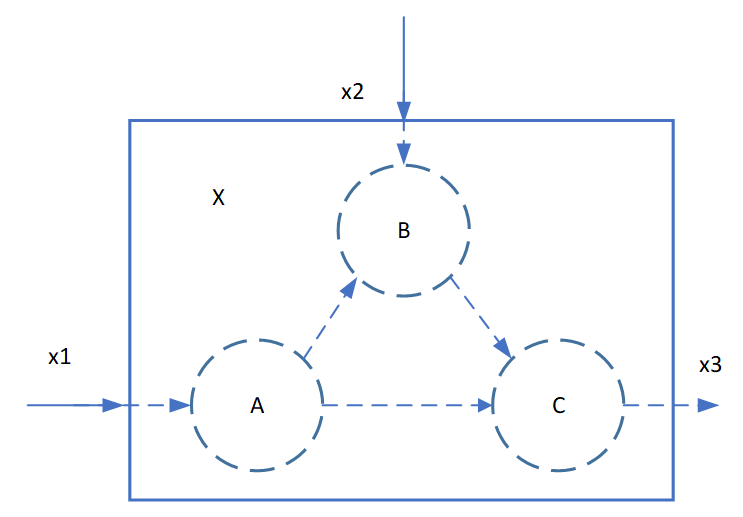}
\caption{\label{fig:f4}System X composed of components A,B,C}
\end{figure}

For example, if the system composed of components A,B,C are regarded as a whole X. Then A,B,C become the internal structure of X, and its mathematical form is also a differential algebraic equation, which is not different from the mathematical form of components A,B,C. The only difference is the number of equations. System X can be treated as a component which has three external interfaces: input interface x1, x2 and output interface x3, as the Figure\ref{fig:f4} shows.

Now, the system and subsystem have a complete concept of closure. If A,B,C are not a minimum system at this time, it can be further subdivided into smaller components. A,B,C can be called a subsystem of system X, and system A,B,C also has its own subsystem.

If A,B,C have only a few equations in total, then A,B,C are not required to be constructed as a subsystem of X. We directly flatten the equation inside X manually. But if A,B,C have thousands of equations, and they also have their own subsystems. Is flattening them manually still a good way?

Therefore, the problems that are divided into components is of great significance when build a giant system. Once the most critical internal structure are determined, people can concentrate on implementing the system (subsystem) at a high level of abstraction.

The rest of the simplification work is left to the computer! If people constantly flatten the equation and build the system from scratch, wouldn't it be a waste of human resources? The value of human thinking should be reflected in the design of the system, not in the flattening of the equation.

\section{Conclusion}

The component-base model has its own advantages and has a broader application prospect. The callback system of ModelingToolkit and the solver of DifferentialEquations can handle simulation problems more flexibly. OptControl.jl\cite{yang2022optcontrol} based on ModelingToolkit can transform the component model system into an optimal control problem solved by JuMP. In addition, there is SciML's\footnote{https://sciml.ai/} parameter identification toolkit and so on. With the support of the Julia community, a component-base model is not only equations for simulation, but it can also be a state space model in control problems, optimizing constraints in problems. Because the essence of components - equations can be described by the most basic Julia element - functions.

\bibliographystyle{unsrt}
\bibliography{sample}

\end{document}